# Observation of Effective Pseudospin Scattering in ZrSiS

*Michael S. Lodge[1,2], Guoqing Chang[3,4], Cheng-Yi Huang[5], Bahadur Singh[3,4], Jack Hellerstedt[6,7], Mark T. Edmonds[6,7,8], Dariusz Kaczorowski[9], Md Mofazzel Hosen[1], Madhab Neupane[1], Hsin Lin[3,4], Michael S. Fuhrer[6,7,8], Bent Weber[6,7,8,10,\*], and Masahiro Ishigami[1,2]*

[1] Department of Physics, University of Central Florida, Orlando, Florida 32816, USA

[2] NanoScience Technology Center, University of Central Florida, Orlando, Florida 32826, USA

[3] Centre for Advanced 2D Materials and Graphene Research Centre National University of Singapore, Singapore 117546

[4] Department of Physics, National University of Singapore, Singapore 117542

[5] Institute of Physics, Academia Sinica, Taipei 11529, Taiwan

[6] Monash Centre for Atomically Thin Materials, Monash University, Clayton VIC 3800, Australia

[7] School of Physics and Astronomy, Monash University, Clayton VIC 3800, Australia

[8] ARC Center of Excellence in Future Low-Energy Electronics Technologies, Monash University, Victoria 3800, Australia

[9] Institute of Low Temperature and Structure Research,

Polish Academy of Sciences, 50-950 Wroclaw, Poland

[10] School of Physical and Mathematical Sciences, Nanyang Technological University,

Singapore 637371

*Corresponding Author Email: b.weber@ntu.edu.sg; Phone: +65 6904 1249

ABSTRACT: 3D Dirac semimetals are an emerging class of materials that possess topological electronic states with a Dirac dispersion in their bulk. In nodal-line Dirac semimetals, the conductance and valence bands connect along a closed path in momentum space, leading to the prediction of pseudospin vortex rings and pseudospin skyrmions. Here, we use Fourier transform scanning tunneling spectroscopy (FT-STS) at 4.5 K to resolve quasiparticle interference (QPI) patterns at single defect centers on the surface of the line nodal semimetal zirconium silicon sulfide (ZrSiS). Our QPI measurements show pseudospin conservation at energies close to the line node. In addition, we determine the Fermi velocity to be $\hbar v_F = 2.65 \pm 0.10$ eV Å in the Γ-M direction ~300 meV above the Fermi energy $E_F$, and the line node to be ~140 meV above $E_F$. More importantly, we find that certain scatterers can introduce energy-dependent non-preservation of pseudospins, giving rise to effective scattering between states with opposite pseudospin deep inside valence and conduction bands. Further investigations of quasiparticle interference at the atomic level will aid defect engineering at the synthesis level, needed for the development of lower-power electronics via dissipationless electronic transport in the future.

In contrast to three-dimensional (3D) topological insulators[1,2], which have metallic surface states but are insulators in the bulk, topological semimetals exhibit topological protection of bulk semimetallic electronic states. In nodal line semimetals, the conduction and valence bands meet along a closed path in momentum space, different from the point-like crossings of Dirac and Weyl semimetals. Recent observations of large, anisotropic magnetoresistance[3-6] have been attributed to resonant electron-hole compensation and orbital topology at the Fermi surface[3], making this material attractive for geomagnetism sensing applications. Furthermore, the presence of line nodes can lead to novel physical phenomena such as spin vortex rings and skyrmionic pseudospin patterns[7], a theoretically predicted "maximal" anomalous Hall effect[7], and optically tunable topological phase transitions[8].

Early realizations of nodal line semimetals include $PbTaSe_2$[9] and $PtSn_4$[10]. However, additional electronic states at energy comparable to the nodal line have been shown to disrupt the emergent topological transport properties in these materials[11], making effective leveraging of the unique properties difficult. More recently, nodal line semimetals without disruptive vicinal bands have been realized, such as the ZrSiX compounds[12-15], where X = S, Se, and Te.

ZrSiS, zirconium silicon sulfide, has been predicted to host a closed Dirac line node protected by glide mirror symmetry of the lattice[16]. This material is particularly interesting because its line node is predicted to exist near the Fermi energy with linear dispersion over a wide energy range unlike other nodal-line materials. Previous angle resolved photoemission spectroscopy (ARPES) studies have confirmed a linear dispersion and closed-loop Fermi surface in the occupied states[12,13]. ARPES results also revealed an additional Dirac surface state, protected by non-symmorphic symmetry[12], below the Fermi energy. However, ARPES measurements, sensitive to occupied states below $E_F$, could not visualize the line node directly. The position of the line node was inferred to be 150-320 meV above $E_F$ from a linear extrapolation of the occupied bands[12,13]. Complementary experimental techniques are therefore necessary to characterize the unoccupied states and observe electronic states near the line node.

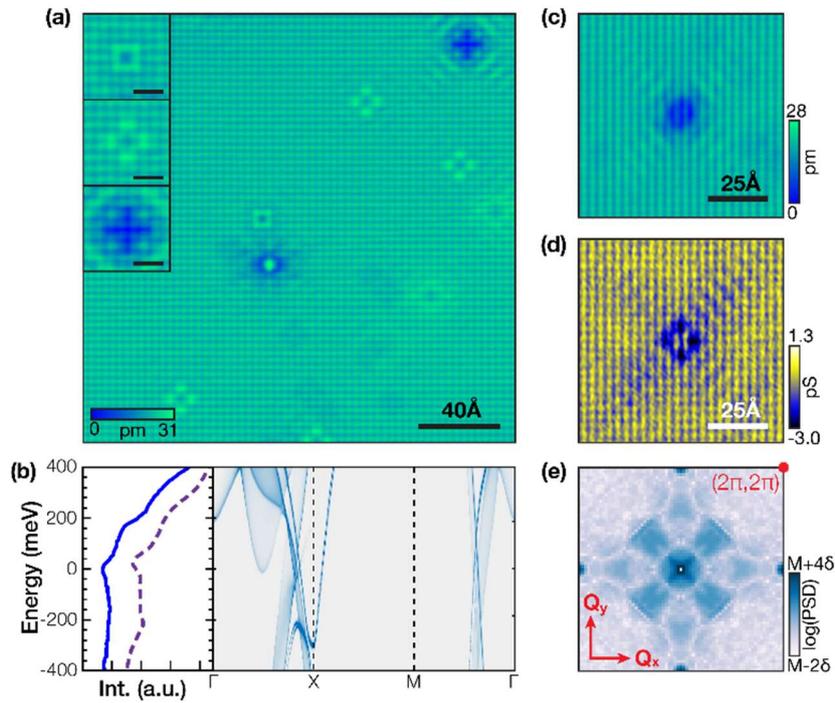

**Figure 1.** (a) STM topograph acquired at T = 4.5K of the as-cleaved ZrSiS surface illustrating a variety of single-atom lattice defects. (Insets) Higher magnification topographic images of three subsurface defect types. The scale bars are each 1 nm. (b) Scanning tunneling spectrum (blue) and the calculated LDOS (dashed purple) of pristine ZrSiS are shown on the left side of the panel. The calculated surface spectral weight of ZrSiS is shown on the right side of the panel for comparison. (c) STM topograph of an individual, single-atom defect. (d) $dI/dV$ map in the same area as (c) acquired at +150 meV. (e) Rotation and reflection averaged 2D fast Fourier transform of (d). The crystal directions $Q_x$ and $Q_y$ are indicated.

In this paper, we use Fourier transform scanning tunneling microscopy (FT-STM) at T = 4.5 K to observe quasiparticle interference (QPI) in both the occupied and unoccupied states of ZrSiS arising from scattering and interactions with individual single-atom defects. We identify the scattering vectors giving rise to our observed constant-energy QPI patterns, which allows us to determine the line node energy and position in reciprocal space for the Γ-M crystal direction, as well as the Fermi velocity in the conduction band. A detailed comparison of the experimental results with theoretical models suggests that some point scatterers can suppress topological protection against backscattering at energies away from the line node.

Single crystals of ZrSiS were synthesized via vapor phase transport methods described elsewhere[6, 17] and samples cleaved in UHV at room temperature by removing a ceramic post that had been epoxied ex situ to the surface of the samples. Cleaving was repeated several times in the same conditions and revealed no variations in the as-cleaved surface topography. STM and STS measurements were performed in a commercial Createc LT-STM at 4.5 K with a platinum iridium tip. QPI data was acquired by spatially mapping the dI/dV at fixed tunneling biases in sequence of decreasing tunneling bias. This method mitigates the detrimental effects of lateral tip drift inherent in commercial LT-STM systems when performing the fast Fourier transform (FFT). Topography and differential conductivity mapping were done in constant current mode. Spectroscopy and differential conductivity mapping measurements were performed using a lock-in amplifier at a signal modulation frequency of 707 Hz with an amplitude of 20 mV. Scanning tunneling spectroscopy measurements were calibrated against the Shockley surface state of Au(111).

An atomically resolved STM image of the ZrSiS surface is shown in Fig. 1a. ZrSiS is a layered compound of tetragonal crystal structure belonging to the P4/nmm space group. A planar square lattice of Si atoms is sandwiched between ZrS bilayers, each of which adopts square lattice geometry. The material is expected to cleave between adjacent ZrS layers. This is consistent with our atomic-resolution image, revealing large, atomically-flat terraces and a lattice constant $a$ = 3.54 ± 0.16 Å, suggesting S termination. A variety of atomic-scale defects show standing wave patterns in the local density of states (LDOS), arising from interference of scattered quasiparticles (inset, Fig. 1a).

Fig. 1b shows the comparison of our tunneling spectroscopy data (STS), acquired in a region unaffected by defects, along with the theoretical local density of states (LDOS) from the top S and Zr atoms. STS data acquired directly over an identical defect shows no major variations from STS acquired far from the defect center. The electronic structure was calculated using the projector augmented wave (PAW) method within the density functional theory (DFT)[18] framework as implemented in the Vienna ab initio simulation package (VASP).[19,20] The generalized gradient approximation (GGA) was used to include the exchange-correlation effects.[21] Spin-orbit coupling (SOC) was included self-consistently. In order to calculate surface states, we generated Wannier functions for both bulk ZrSiS and its 12 atom layers' thin film, using Zr d, Si s and p and S p orbitals.[22] The calculated LDOS is found to be shifted by +45 meV relative to our experimental data, but otherwise is in a good agreement with experiment; such shift likely reflects doping due to atomic defects such as vacancies or substituents. The most prominent feature in both spectra is the slight dip in the differential conductance near 0 meV, which has previously been ascribed to the Dirac line node.[23] The right side of Fig. 1b shows the full theoretical surface spectral weight (SSW) spectrum, which shows bulk and surface density of states resolved in energy and wave vector in four different crystal directions, similar to a band structure. Pure surface states and surface resonances appear in the SSW as dark, sharp lines, and projected bulk states appear as broad, diffuse regions. Many overlapping surface and bulk states near the X point between 25 meV< $E$ <75 meV prevent clear attribution of the dip feature in the STS the line node. Therefore, tunneling point spectroscopy alone is insufficient to clearly identify the line node.

To gain more detailed information of the ZrSiS electronic band structure, we turn to QPI measurements. Fig. 1c shows an STM topograph of a single subsurface defect, measured at a sample bias of +150 mV. All QPI data included in this manuscript has been taken in the vicinity of this particular defect. Fig. 1d shows the corresponding differential conductance measurement acquired simultaneously with the topography. Fig. 1e shows a 2D Fourier transform of the same area reflecting the fourfold symmetry of the crystal as seen in the conductance map of Fig. 1d. For a better signal-to-noise ratio, the QPI pattern as shown has been symmetrized by rotation and reflection averaging along high symmetry directions, taking the fourfold symmetry of the crystal into account.

We now show how the features of the QPI pattern arise as a direct consequence of the shape of the SSW. Fig. 2a shows the SSW at -100 meV. For ZrSiS, the SSW in the vicinity of the line node takes the shape of two concentric squares that make up the opposing branches of the Dirac dispersion. Combined with scattering selection rules, the SSW governs the scattering wave vectors $Q = k_f - k_i$ (red arrows) for quasiparticle scattering between initial (*i*) and final (*f*) states.

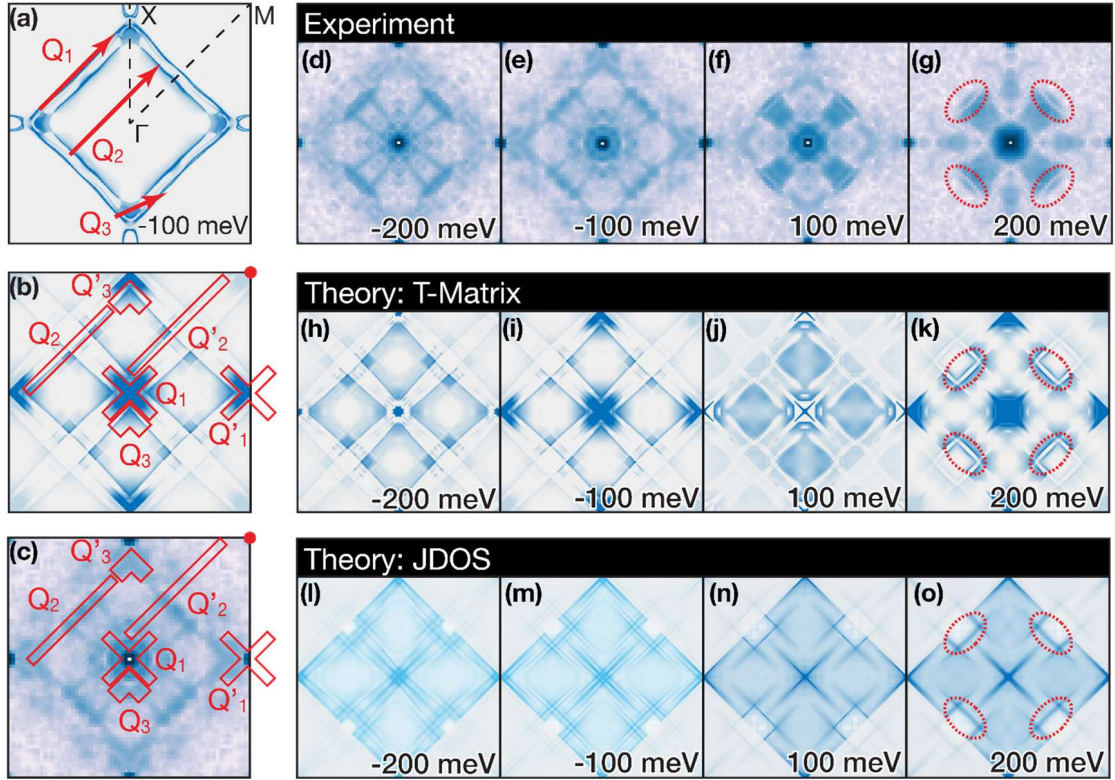

**Figure 2.** (a) Isoenergentic contour of the calculated surface spectral weight at -100 meV illustrating possible nesting vectors. (b) T-Matrix calculation of the QPI pattern generated from (a). Outlined in red are the features corresponding to the vectors $Q_1$-$Q_3$ shown in (a). Features $Q'_1$-$Q'_3$ arise from respective scattering processes between neighboring Brillouin zones. (c) Experimental QPI pattern at the same corresponding energy as (b), with features outlined as in (b). (d-g) Experimental QPI patterns in the vicinity of the line node. (h-k) T-Matrix calculation of the QPI patterns. (l-o) JDOS calculations of the QPI patterns. Red ellipses in (g,k,o) highlight the splitting of the QPI pattern as discussed in text.

We use two theoretical models to understand the impact of selection rules for scattering. The simpler joint density of states (JDOS) model weighs every possible scattering wave vector equally with no consideration given to spin-, pseudospin-, or orbital-selection rules. However, in the real material, we expect certain scattering processes to be suppressed by the selection rules. As such, we also consider the T-matrix method, which takes spin and orbital composition into account, as well as complex matrix element and impurity structure, which can give rise to quantum interference effects absent in JDOS[24]. Both computational methods serve as extreme limits for scattering in this material, as the JDOS represents maximal mixing and the T-matrix (for *T*=1) represents perfect preservation of quantum degrees of freedom. Nevertheless, both approaches are instructional references in the interpretation of our data.

Fig. 2b shows the QPI pattern from the T-matrix calculation corresponding to the SSW in Fig. 2a, and Fig. 2c shows the experimental QPI pattern. We identify six groups of scattering vectors that give rise to the calculated interference pattern, shown schematically in Figs. 2a-c. Scattering vectors $Q_1$ to $Q_3$ denote quasiparticle scattering within the first Brillouin zone, while the scattering vectors labeled $Q_1'$ to $Q_3'$ correspond to similar processes connecting features between adjacent Brillouin zones. We observe all scattering vectors clearly in the measured QPI pattern (Fig. 2c) except for $Q_3$ whose shape is partially obscured by instrumental (low frequency) noise at small wavenumbers. Higher wavenumber features beyond the first Brillouin zone are visible in the theoretical calculations, but are absent in the measurement. This is likely due to a suppression of scattering with large wave vectors.

Figs. 2d-g show the evolution of the measured QPI pattern at energies around the Fermi level. We compare this data with T-matrix (Figs. 2h-k) and JDOS (Figs. 2l-o) models. In our data, we observe quasiparticle interference as sharp lines of finite intensity and width below the Fermi level. Above the Fermi level, we observe similar features, but with significant scattering intensity from the region within the cross shape. The contrasting intensities of the QPI pattern above and below the Fermi level can be understood by considering the *E* and *k* positions of the surface states and the projected bulk states (compare modeled SSW in Fig. 1b). Below the Fermi level, scattering occurs primarily between surface states, resulting in sharp lines of finite width. Above the Fermi level, broad (in energy and *k*-space) bulk states are available for scattering (especially towards the Γ-point of the Brillouin zone), leading to finite intensity within the cross.

We focus on the strong QPI feature produced by the heavily nested set of scattering wave vectors $Q_2$. This feature is strongly dependent on the topological nature of the line node, as pseudospin conservation forbids two of the four possible scattering processes between the two Dirac bands on opposite sides of the Brillouin zone; only scattering from inner to inner and outer to outer bands are allowed. Thus, the T-matrix calculation, which includes pseudopspin conservation, shows a split feature at $Q_2$ (see e.g. red dotted ellipses in Fig. 2k) while the splitting is not evident in JDOS where all four scattering processes are present (red dotted ellipses in Fig. 2o). In the experiment (Fig. 2g) we observe the splitting of the $Q_2$ feature (red ellipses in Fig. 2g) in the unoccupied states (200 meV) – a feature expected for energies at, and immediately above and below, the line node, and is a direct observation of the its topological nature. As such, our QPI measurements are consistent with pseudospin conservation at these energies[25].

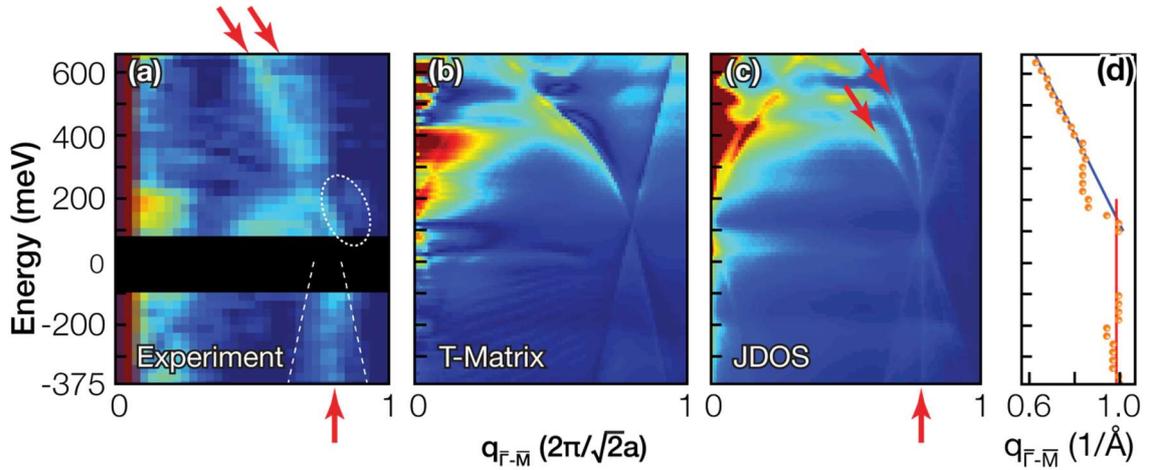

**Figure 3.** (a) The experimentally derived dispersion of the QPI pattern in the Γ-M direction. Red arrows at the top of the panel show two dispersing features above $E_F$, while the red arrow below $E_F$ shows a nondispersing feature. The white, dashed ellipse highlights the scattering suppression in the vicinity of the line node. Dashed white lines show the expected dispersion below $E_F$ from previous ARPES measurements. (b) T-Matrix calculation of the QPI dispersion in the Γ-M direction. (c) JDOS calculation of the QPI dispersion in the same direction as (b). Red arrows in (c) highlight similar features corresponding to observations in (a). (d) Position of the maxima of the intense dispersing feature above ~300 meV and the nondispersing feature below $E_F$ in (a). Lines are linear fits to the data above 300 meV and below -100 meV.

We now look at our data at a larger energy range. Fig. 3 shows the energy dispersion of the QPI along the Γ-M direction for experiment (a), as well as T-matrix (b) and JDOS (c) calculations. We operate the STM at tunneling biases at or outside +/-100 meV of $E_F$ to perform the dI/dV imaging necessary to collect the raw data using the constant current mode. We chose the constant current mode rather than the constant height mode because it produces QPI data with better stability than constant height mode. The splitting observed in Fig. 2g manifests itself in the dispersion (white dotted ellipse in Fig. 3a) up to ~250 meV, showing that pseudospin is conserved up to this energy. At different energies, the data diverges from the T-matrix calculation. We observe two separate branches with negative slope above $E_F$ at large positive energy, and a strong non-dispersive feature, accompanied by a weaker dispersive feature, below $E_F$ (red arrows). The same non-dispersing $q$-vector in the valence was observed in data obtained by both constant current and constant height mode measurements, excluding the set point effect[26] as a possible cause. Both the second negative-slope feature above $E_F$ and the non-dispersive feature below $E_F$ are not present in our T-matrix calculation (Fig. 3b) but are reproduced well in the JDOS calculation (Fig. 3c); both features result from pseudospin non-conserving processes at $Q_2$. Therefore, the data in Fig. 3 show that the investigated defect poses a non-trivial scattering potential for quasiparticles, leading to a partial lifting the expected topological protection at some energy away from the line node. Impurities such as the one investigated will cause additional scattering and limit the quasiparticle mean free path in transport. Importantly, our study shows that useful QPI spectra may be obtained for individual defects of a certain type. We expect that the investigation of the quasiparticle scattering and interferences at different single-atomic defect centers will consequently be an important characterization tool for topological materials, that can be supported by T-matrix calculations with non-trivial scattering potentials.

Finally, we determine the location of the line node and the value of the Fermi velocities. Fig. 3d shows the position of the local maximum of the QPI intensity as a function of energy. Fitting a line to the region $E > 300$ meV reveals a Fermi velocity $\hbar v_F = 2.65 \pm 0.10$ eV-Å, significantly lower than the valence band velocity $\hbar v_F = 7.1$ eV-Å seen in ARPES for the Γ-M direction[13]. The lower velocity results from the deviation from linearity of at least the inner branch of the line node above $E_F$, which is also seen in the SSW calculation. The non-dispersive feature observed in the occupied states is precisely located at the $q$ position of the line node according to the JDOS calculations as shown in Fig. 3c, indicating the line node to be located at $q = 0.98 \pm 0.05$ Å$^{-1}$. Extrapolating the dispersive conduction band feature to this q position results in the line node energy of $140 \pm 40$ meV. This is an overestimate due to the negative curvature of the feature, but it is in reasonable agreement with the position (+45 meV) inferred from the LDOS (Fig. 1b).

In conclusion, we have used dI/dV mapping to image quasiparticle interference from scattering by single defects on the surface of line-nodal semimetal ZrSiS. The Fermi velocity is $\hbar v_F$ = 2.65 ± 0.10 eV Å in Γ-M direction above ~300 meV and the nodal line position is found to be 140 ± 40 meV above $E_F$ at $q$ = 0.98 ± 0.05 Å$^{-1}$. Comparing with ab initio calculations, we have identified six groups of scattering vectors, reflecting the material's Fermi surface in the unoccupied states and providing deeper insight into the material's charge transport properties. Our analysis indicates that the type of atomic defect studied here, provides a mechanism for pseudospin scattering since they couple both pseudospin polarities locally. Further refinement of the computation method beyond the Born limit and beyond T=1 (including off-diagonal and complex elements) may ultimately provide deeper insight into the exact nature of the scattering potential as well as the energy dependence of the pseudospin-flip scattering observed.[24,25] Further experiments may allow us to disentangle the role of individual atomic defect centers in quasiparticle scattering, and determine their impact on topological protection of electronic states.

Since submission of this manuscript, we became aware of another work based on QPI observations at the ZrSiS surface[27].

AUTHOR INFORMATION

**Corresponding Author**

*Email: b.weber@ntu.edu.sg**Author Contributions**

MSL and BW, along with MI, MN and MSF conceived the experiment; DK obtained the single crystals used in the STS experiments and proved their high-quality via bulk crystal-chemical and physical measurements; MMH discussed the sample preparation with MSL; JH and ME assisted in developing the experimental apparatus at Monash; MSL and MI performed preliminary imaging experiments at UCF; MSL and BW performed the STM and QPI measurements at Monash; MSL and BW analyzed the experimental data with input from MI and MSF; GC, CYH, BS, and HL calculated the theoretical QPI data; MSL and BW prepared the manuscript with input from the other authors. All authors have seen the data and agreed upon its interpretation.

ACKNOWLEDGEMENT

MSL was supported by the East Asia and Pacific Summer Institute of the National Science Foundation and the Australian Academy of Sciences. This material is also based upon work supported by the National Science Foundation under Grant no. 0955625 (MSL and MI). BW acknowledges an Australian Research Council DECRA fellowship and a Singapore National Research Foundation (NRF) Fellowship.

SUPPORTING INFORMATION

Comparison of closed feedback vs. open feedback quasiparticle interference data acquisition.

For TOC only:

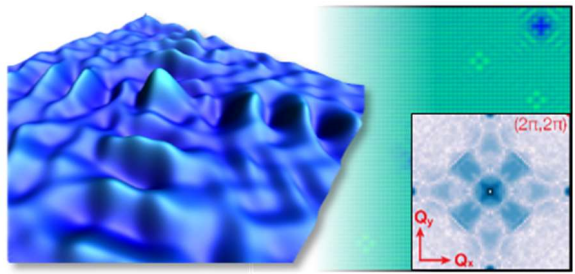